\documentclass[a4paper]{ESLAB2005}
\usepackage{epsfig,amssymb}

\begin{document}

\title{A Mission to Explore the Pioneer Anomaly}

\author{
H.~Dittus,\inst{1} S.\,G.~Turyshev,\inst{2} 
C.~L\"ammerzahl,\inst{1} S.~Theil,\inst{1}
R.~Foerstner,\inst{3} U.~Johann,\inst{3}
W.~Ertmer,\inst{4} E.~Rasel,\inst{4}
B.~Dachwald,\inst{5} W.~Seboldt,\inst{5}
F.\,W.~Hehl,\inst{6,21} C.~Kiefer,\inst{6}
H.-J.~Blome,\inst{7}
J.~Kunz,\inst{8}
D.~Giulini,\inst{9}      
R.~Bingham,\inst{10} B.~Kent,\inst{10}
T.\,J.~Sumner,\inst{11}
O.~Bertolami,\inst{12} J.~P\'aramos,\inst{12}
J.\,L.~Rosales,\inst{13}
B.~Christophe,\inst{14} B.~Foulon,\inst{14} P.~Touboul,\inst{14}
P.~Bouyer,\inst{15} 
S.~Reynaud,\inst{16}
A.~Brillet,\inst{17} F.~Bondu,\inst{17} E.~Samain,\inst{17} 
C.\,J.~de~Matos,\inst{18}
C.~Erd,\inst{19} J.\,C.~Grenouilleau,\inst{19} D.~Izzo,\inst{19} A.~Rathke,\inst{19}
\\
J.\,D.~Anderson,\inst{2} S.\,W.~Asmar,\inst{2} E.\,E.~Lau,\inst{2} 
M.\,M.~Nieto,\inst{20}
\and 
B.~Mashhoon\inst{21}} 
\institute{
Centre of Applied Space Technology \& Microgravity (ZARM), 
University of Bremen, Am Fallturm, 
28359 Bremen, 
Germany
\and 
Jet Propulsion Laboratory, 
California Institute of Technology, 
4800 Oak Grove Drive, Pasadena, CA 91109, USA 
\and 
Department of Science Programs, 
Earth Observation and Science,
Astrium GmbH, 
88039 Friedrichshafen,
Germany
\and 
Institute for Quantum Optics, University of Hannover
Welfengarten 1,
30167 Hannover,
Germany 
\and 
German Aerospace Center (DLR), 
Institute of Space Simulation,
Linder Hoehe, D-51170 K\"oln, Germany
\and
Institute for Theoretical Physics,
University of Cologne,
50923 K\"oln, Germany
\and
University of Applied Sciences,
Department of Space Technology,
Hohenstaufenallee 6,
52064 Aachen,
Germany
\and 
Institute for Physics,
Carl von Ossietzki University Oldenburg,
26111 Oldenburg,
Germany
\and
Institute of Physics,
Albert-Ludwigs-University,
Hermann-Herder-Strasse 3,
D-79104 Freiburg im Breisgau,
Germany
\and
Space Engineering and Technology Division,
Rutherford Appleton Laboratory,
Chilton,
Oxfordshire OX11 0QX,
UK
\and 
Imperial College, 
The Blackett Laboratory, 
Prince Consort Road, London SW7 2BZ, 
London, UK
\and 
Instituto Superior T\'ecnico, Departamento de F\'isica,  
Av. Rovisco Pais, 1, 1049-001 Lisboa, Portugal
\and 
Quantum Information Group, RSFE, and 
Xerox Corporation,
S.A.U. Ribera del Loira 16-18,
28042 Madrid,
Spain
\and 
Physics \& Instrumentation Department, 
ONERA,
BP72,
29 ave de la division Leclerc,
92322 Chatillon,
France
\and 
Laboratoire Charles Fabry de l'Institut d'Optique,
Bat 503,
Centre Scientifique,
91403 Orsay,
France
\and 
Laboratoire Kastler Brossel,
Universit\'e Pierre et Marie Curie,
Campus Jussieu, case 74,
75252 Paris,
France
\and 
Observatoire de la C\^ote d'Azur, 
CERGA, Av. N. Copernic, 
F-06130 Grasse, France
\and
Advanced Concepts and Studies,
ESA-HQ (SER-A),
8, 10 Rue Mario Nikis,
75738 Paris,
France
\and 
ESA-ESTEC, 
Keplerlaan 1,
2201 AZ Noordwijk ZH,
The Netherlands	
\and  
Theoretical Division (MS-B285), 
Los Alamos National Laboratory,
University of California, Los Alamos, NM 87545, USA
\and 
Department of Physics and Astronomy, University of Missouri, Columbia, MO 65211, USA 
}

\maketitle 

\begin{abstract}

The Pioneer 10 and 11 spacecraft yielded the most precise navigation in deep space to date. These spacecraft had exceptional acceleration sensitivity. However, analysis of their radio-metric tracking data has consistently indicated that at heliocentric distances of $\sim$20--70 astronomical units, the orbit determinations indicated the presence of a small, anomalous, Doppler frequency drift. The drift is a blue-shift, uniformly changing with a rate of $\sim(5.99 \pm 0.01)\times 10^{-9}$~Hz/s, which can be interpreted as a constant sunward acceleration of each particular spacecraft of $a_P  = (8.74 \pm 1.33)\times 10^{-10}$~m/s$^2$ (\cite{pioprl,moriond,pioprd}). The nature of this anomaly remains unexplained. This signal has become known as the Pioneer anomaly.

The inability to explain the anomalous behavior of the Pioneers with conventional physics has contributed to growing discussion about its origin. There is now an increasing number of proposals that attempt to explain the anomaly outside conventional physics. This progress emphasizes the need for a new experiment to explore the detected signal. Furthermore, the recent extensive efforts led to the conclusion that only a dedicated experiment could ultimately determine the nature of the found signal.

We discuss the Pioneer anomaly and present the next steps towards an understanding of its origin.  We specifically focus on the development of a mission to explore the Pioneer Anomaly in a dedicated experiment conducted in deep space.  This joint European-US mission is motivated by the desire to better understand the laws of fundamental physics as they affect dynamics in the solar system. The mission could lead to a major discovery in the 21st century and, with readily available technologies, it could be flown well within the Cosmic Vision time frame.

\end{abstract}

\section{Background}
\label{intro}

The exploration of the solar system's frontiers - the region between 25-250 astronomical units (AU) from the Sun - is a most ambitious and exciting technological challenge.  The scientific goals for possible deep-space missions are well-recognized and include studies of the gas and dust distributions, exploration of the heliopause and the space beyond, measurements of the magnetic fields and particle fluxes, studies of the Oort Cloud and Kuiper Belt Objects, encounters with distant bodies, and investigation of the dynamical background of the solar system by studying the effects of various forces that influence the trajectory of the spacecraft.  We are most interested in this last goal.  

Our interest comes from navigating the Pioneer 10 and 11 spacecraft that yielded an exceptionally good acceleration sensitivity.  Surprisingly, the accuracies of their orbit reconstruction were limited by a small, anomalous, Doppler frequency drift that can be interpreted as a sunward constant acceleration of the craft of $a_P  = (8.74 \pm 1.33)\times 10^{-10}$~m/s$^2$ (see \cite{pioprd}).  This interpretation has become known as the Pioneer anomaly.

The nature of this anomaly remains a mystery, with possible explanations ranging from nominal sources of on-board systematics to exotic gravity extensions on solar system scales. Although the most obvious cause would be that there is a systematic origin to the effect, the limited data analyzed does not unambiguously support any of the suggested mechanisms (\cite{pioprd}). The inability either to explain the anomaly or to test it with other spacecraft has contributed to a growing discussion about its origin (\cite{cospar,pio-mission,mex,stanford}). 

Recently there was a significant interest in developing a dedicated mission to study the detected signal.  Previous extensive efforts have included formulating theoretical mechanisms to explain the anomaly and analyzing existing solar system data, including both planetary and spacecraft data.  Analysis of the capabilities of spacecraft currently in operation or in design demonstrated their inability to fulfill an independent verification of the anomaly.  These efforts led to the conclusion that only a dedicated experiment could ultimately determine the nature of the anomalous signal.  

The paper is organized as follows: In Section \ref{mission} we discuss the Pioneer missions and the detected anomaly. We also review   mechanisms proposed to explain the Pioneer anomaly, both with conventional and `new' physics. Section \ref{missions} discusses the program of experimental tests, focusing on a dedicated mission concept to explore the Pioneer anomaly, and in Section \ref{conclude} we present our conclusions.


\section{The Pioneer Missions and the Anomaly}
\label{mission}

\begin{figure}[t!]
\centering \vskip -8pt 
\epsfig{figure=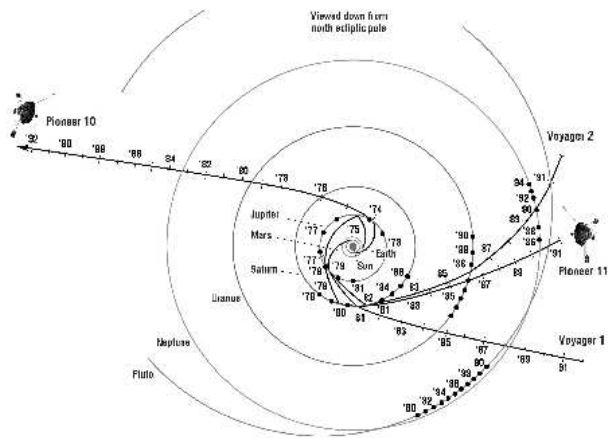,angle=-90,width=9.1cm}
\caption
{Ecliptic pole view of Pioneer and Voyager trajectories.  Pioneer 11 is traveling approximately in the direction of the Sun's orbital motion about the galactic center.  The galactic center 
is approximately in the direction of the top of the figure.} 
     \label{fig:pioneer_path}
\end{figure}

The Pioneer 10/11 missions, launched on 2 March 1972 (Pioneer
10) and 5 April 1973 (Pioneer 11), respectively, were the first spacecraft to explore the outer solar system (\cite{pioprd}). After Jupiter and (for Pioneer 11) Saturn encounters, the craft followed escape hyperbolic orbits near the plane of the ecliptic to opposite sides of the solar system. (See Figure \ref{fig:pioneer_path}.)
Pioneer 10 eventually became the first man-made object to leave the solar system.  The last telemetry was obtained from Pioneer 10 on 27 Apr 2002 when the craft was 80 AU from the Sun. (The last signal from Pioneer 10 was received on 23 Jan 2003.)    

\begin{figure}[h!]
 \begin{center}
\noindent \vskip-14pt   
\psfig{figure=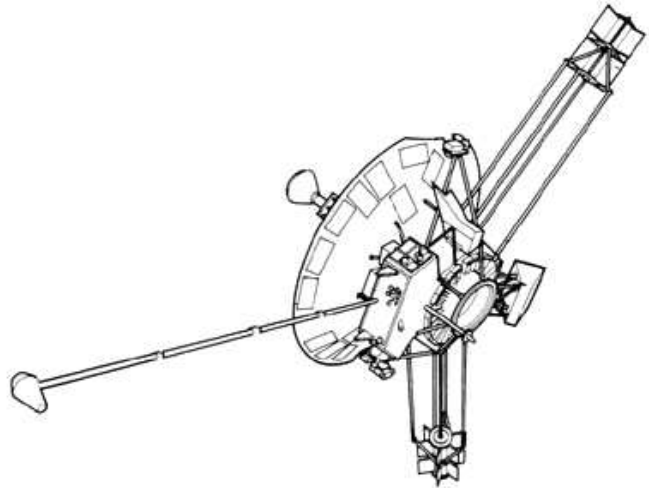,angle=-90,width=9.1cm}
\end{center}
  \caption{A drawing of the Pioneer spacecraft.  
 \label{fig:craft}}
\vskip -8pt 
\end{figure} 

The Pioneers were excellent craft with which to perform precise celestial mechanics experiments.  This was due to a combination of many factors, including their attitude control (spin-stabilized, with a minimum number of attitude correction maneuvers using thrusters), power design (the Plutonium-238 powered heat-source RTGs -- Radioisotope Thermoelectric Generators -- being on extended booms aided the stability of the craft and also reduced the effects due to heating), and precise Doppler tracking (with the accuracy of post-fit Doppler residuals at the level of mHz).  The result was the most precise navigation in deep space to date. (See Fig.~\ref{fig:craft} for a design drawing of the spacecraft.)

By 1980, when Pioneer 10 passed a distance of  $\sim$ 20 AU from the Sun, the acceleration contribution from solar-radiation pressure on the craft (directed away from the Sun) had decreased to less than $4
\times 10^{-10}$ m/s$^2$.   This meant that small effects could unambiguously be determined from the data, and the anomalous acceleration began to be seen.  A detailed study of the anomaly began in 1994, using data starting in 1987.0. By then the  external systematics (like solar-radiation pressure) were limited and the existence of the anomaly in the Pioneers' data became clearly evident (\cite{pioprd,mex,cospar,stanford}). 

In the next section, we shall review our current knowledge of the Pioneer anomaly. 

\subsection{A summary of the Pioneer Anomaly}
\label{anomaly}

As discussed above, the analysis of the Pioneer 10 and 11 data (\cite{pioprd}) demonstrated the presence of an anomalous, Doppler frequency blue-shift drift, uniformly changing with a rate of (\cite{stanford})
\begin{equation}
\dot{f}_P \sim (5.99 \pm 0.01)\times 10^{-9} ~~ \mathrm{Hz/s}.
\end{equation}

To understand the phenomenology of the effect, consider  $f_{\tt obs}$, the frequency of the re-transmitted signal observed by a DSN antenna, and $f_{\tt model}$,  the predicted frequency  of that signal. The observed, two-way (round-trip) anomalous effect can be expressed to first order in $v/c$ as  
\begin{eqnarray}
\left[f_{\tt obs}(t)- f_{\tt model}(t)\right]_{\tt DSN}
=  -2\dot{f}_P\,t, 
\label{eq:delta_nu}
\end{eqnarray}
with  $f_{\tt model}$ being the modeled frequency change due to conventional forces accounted for in the spacecraft's motion.  (For more details see \cite*{moriond,pioprd}.) This motion is outwards from the Sun and hence it produces a red shift.

After accounting for the gravitational and other large forces included in standard orbit determination programs this translates to 
\begin{eqnarray}
\left[f_{\tt obs}(t)- f_{\tt model}(t)\right]_{\tt DSN}
= -f_{0}\frac{2a_P~t}{c}. 
\label{eq:delta_nu_syst}
\end{eqnarray}
Here $f_{0}$ is the reference frequency (\cite{pioprd}). 
 
Furthermore, after accounting for all {\it known} (not modeled) sources of systematic error (discussed in \cite{pioprd}), the conclusion remained that there was an anomalous sunward constant acceleration signal of 
\begin{equation}
a_P=(8.74\pm1.33)\times 10^{-10}~~{\rm m/s}^2. 
\end{equation}
We have already included the sign showing that $a_P$ is inward using the DSN convention. (See \cite*{pioprd} for more information.) 

The initial results of the study were reported in (\cite{pioprl,moriond}) and a detailed analysis appeared in (\cite{pioprd}).  Realizing the potential significance of the result, all {\it known} sources of a possible systematic origin for the detected anomaly were specifically addressed. We emphasize {\it known} because one might naturally expect that there is a systematic origin to the effect, perhaps generated by the spacecraft themselves from excessive heat or propulsion gas leaks.  However, a convincing explanation for the anomalous behavior of the  spacecraft is still unavailable (for more details see discussion in \cite{pioprl,pioprd,old4}).

For the most detailed analysis of the Pioneer anomaly to date, \cite*{pioprd} used the following Pioneer 10/11 Doppler data (\cite{stanford}): 
\begin{itemize}
\item Pioneer 10: The data used was obtained between 3 January 1987 and 22 July 1998. This interval covers heliocentric distances  $\sim40-70.5$~AU. This data set had 20,055 data points obtained over the 11.5 years. 
\item Pioneer 11: The data used was obtained between 5 January 1987 to 1 October 1990. This interval covers heliocentric distances $\sim22.42-31.7$~AU. This data set had 10,616 data points obtained over the 3.75 years. 
\end{itemize}

By now several studies of the Pioneer Doppler navigational data have demonstrated that the anomaly is unambiguously present in the Pioneer 10 and 11 data. These studies were performed with three independent (and different!) navigational computer programs 
(\cite{pioprl,pioprd,markwardt}).  Namely: 
{}
\begin{itemize}
\item Various versions of JPL's Orbit Determination Program (ODP) code developed in 1980-2005,
\item a version of The Aerospace Corporation's  CHASPM (latest version of POEAS, see \cite{pioprd}) code extended for deep space navigation, and finally 
\item a third code written by C.~Markwardt (\cite{markwardt}), of the Goddard Space Flight Center (GFSC).   He analyzed Pioneer 10 data obtained from the National Space Science Data Center (NSSDC, with more information at {\tt http://nssdc.gsfc.nasa.gov/}), for the time period 1987-1994. 
\end{itemize}

Several analyses of the Pioneer 10 and 11 radio-metric data (\cite{pioprl,moriond,pioprd,markwardt,cospar,mex,stanford}) have established the following basic properties of the Pioneer anomaly:
{}
\begin{itemize}
\item {\it Direction:} Within the 10 dbm bandwidth of the Pioneer high-gain antennae, the anomaly behaves as a line-of-sight constant acceleration of the spacecraft directed toward the Sun.
\item {\it Distance:} It is unclear how far out the anomaly goes, but the Pioneer 10 data supports the presence of the anomaly at distances up to $\sim$70~AU from the Sun. In addition, the Pioneer 11 Doppler data shows the presence of the anomalous constant frequency drift as close in as $\sim$20 AU. 
\item {\it Constancy:}~Both temporal and spatial variations of the anomaly's magnitude are less then 3.4\% for each craft.
\end{itemize}

This information was used as guidance in investigating applicability of proposals suggested to explain the Pioneer anomaly with both conventional and `new' physical mechanisms. In Section~\ref{sec:explain} we will briefly review these proposals.  We will use the same principles in Section~\ref{missions} while discussing our proposal to explore the Pioneer anomaly in a dedicated deep space experiment. 

\subsection{Attempts at explanations}
\label{sec:explain}
\subsubsection{Conventional physics mechanisms:} 

There were many attempts to explain the anomaly with a conventional physics mechanism which are either not strong enough to explain its magnitude or else exhibited significant temporal or spatial variations contradicting the known properties of the anomaly. Consequently, attempts of explanation along the lines of conventional physics have addressed a number of possibilities, namely:

\noindent{\it Unknown mass distribution in the solar system.} The most straightforward way to generate a putative physical force is the gravitational attraction due to an imprecisely known mass distribution in the outer solar system. Such a distribution could be due to the Kuiper Belt Objects or dust. The known density distribution for the Kuiper belt has been studied in \cite*{pioprd}, and found to be incompatible with the discovered properties of the anomaly. Even worse, these distributions cannot circumvent the constraint from the undisturbed orbits of Mars and Jupiter (\cite{mmn05,orfeu05}). Hence a gravitational attraction by the Kuiper belt can, to a large extent, be ruled out.

\noindent{\it Interplanetary dust} leads to (i) a gravitational acceleration, (ii) an additional drag force (resistance) and (iii) a frequency shift of the radio signals proportional to the distance. The analysis of data from the inner parts of the solar system taken by the Pioneer 10 and 11 dust detectors strongly favors a spherical distribution of dust over a disk. Ulysses and Galileo measurements in the inner solar system find very few dust grains in the $10^{-18}-10^{-12}$~kg range (\cite{dust}). IR observations rule out more than 0.3 Earth mass from Kuiper Belt dust in the trans-Neptunian region. Furthermore, the density varies greatly within the Kuiper belt, precluding any constant acceleration. The density of dust is not large enough to produce a gravitational acceleration on the order of $a_P$ (\cite{pioprd}). The resistance caused by the interplanetary dust is too small to provide support for the anomaly, so as the dust-induced frequency shift of the carrier signal.

Recently there was a revived interest in examining the dust distribution in the outer solar system as a possible origin of the Pioneer anomaly.  \cite*{dust} considered this idea based on the known properties of dust in the solar system which is composed of a thinly scattered matter with two main contributions:
\begin{itemize}
\item Interplanetary Dust (IPD): a hot-wind plasma (mainly $p$ and $e^-$) within the Kuiper Belt,  from 30 to 100~AU with a modeled density of $\rho_{\rm IPD}\lesssim10^{-21}$~kg/m$^3$;
\item Interstellar Dust (ISD): composed of fractions of interstellar dust (characterized by greater impact velocity). The density of ISD was directly measured by the Ulysses spacecraft, yielding $\rho_{\rm ISD}\lesssim3\times10^{-23}$~kg/m$^3$. 
\end{itemize}
\cite*{dust} used these properties to estimate the effect of dust on the Pioneers and found that one needs an axially-symmetric dust distribution within $20-70$~AU with a constant, uniform, and unreasonably high density of
$\sim3\times 10^{-16}$~kg/m$^3\simeq300,000\cdot(\rho_{\rm IPD}+\rho_{\rm ISD})$. Therefore, dust cannot explain the Pioneer anomaly.

\noindent{\it Spin-rotation coupling.} The spin-rotation coupling on the circularly-polarized radio signal when it interacts with the rotation of a spacecraft and the Earth leads to a constant
acceleration which, however, is too small to explain $a_P$ (\cite{spin-rot}). Furthermore, the helicity-rotation coupling has already been phenomenologically incorporated in the analysis of Doppler data.

\noindent{\it Local effect of expansion of the Universe.} Motivated by the numerical coincidence $a_P\simeq cH_0$ where $c$ the speed of light and $H_0$ is the Hubble constant at its present time there are many attempts to explain the anomaly in terms of the expansion of the Universe. \cite*{pioprd} have shown that such a mechanism would produce an opposite sign for the effect and it has been argued (\cite{ranada}) that the cosmic expansion influences the measurement process via a change in the frequency of the traveling electromagnetic signals. However, one expects that taking all effects of the cosmic expansion on the frequency as well as on the Pioneer motion into account, the resulting acceleration is $-vH$ and, thus, has the correct sign but is too small by a factor $v/c$ (\cite{LaemmerzahlDittus05}). The ways in which the cosmic expansion might be responsible for $a_P$ vary considerably between the approaches. It is known (\cite{pioprl,pioprd}) that the very presence of the Pioneer anomalous acceleration contradicts the accurately known motion of the inner planets of our solar system. This motivated focusing on the effect of cosmic acceleration on the radio communication signal rather than on the spacecraft themselves. This approach avoids confronting the anomaly with the present accuracy of the planetary orbits. 

Another line is whether the cosmic expansion has an influence on the size of the Solar system. In particular, the properties of bound
(either electrically or gravitationally) systems in an expanding universe has been discussed controversially in many papers, notably by (\cite{McVittie,anderson_jl}). Very interesting is
the result (\cite{anderson_jl}) where it has been found that the expansion couples to escape orbits, while it does not couple to bound orbits.

\subsubsection{Possibility for new physics?} 
\label{sec:new-ph}

The apparent difficulty to explain the anomaly within standard physics became a motivation to look for `new physics.' These attempts in general did not produce a viable mechanism for the anomaly study, in particular:

\noindent{\it Extensions of general relativity.} An inverse time dependence for the gravitational constant $G$ produces effects similar to that of an expanding universe. So does a length or momentum scale-dependent cosmological term in the gravitational action functional (\cite{Modanese,Rosales}). The anomalous acceleration could be explained in the frame of a quasi–metric theory of relativity (\cite{Ostvang}). The possible influence of the cosmological constant on the motion of inertial systems leading to an additional acceleration has been discussed by \cite*{Rosales03}. 
In addition, there were ideas to invoke a model for superstrong interaction of photons or massive bodies with the graviton background (\cite{Ivanov}). 
A 5-dimensional cosmological model with a variable extra dimensional scale factor in a static external space (\cite{Belayev}) was also proposed. There is also an attempt to explain the anomaly in the framework of a non–symmetric gravitational theory (\cite{moffat}).

\noindent{\it Gravity modifications.} One approach, called MOdified Newtonian Dynamics (MOND), induces a long-range modification of gravity  that explains the rotation curves of galaxies (\cite{Milgrom,Bekenstein}). It has been pointed out that a scalar field with a suitable potential can account for a constant acceleration as experienced by the spacecraft (\cite{moffat05}). A modification of the gravitational field equations for a metric gravity field, by introducing a general linear relation between the Einstein tensor and the energy-momentum tensor has been claimed to account for $a_P$ (\cite{serge,serge2}).

\noindent{\it Dark matter.} Various distributions of dark matter in the Solar system have been proposed to explain the anomaly, e.g., dark matter distributed in a form of a disk in the outer solar system with of a density of $\sim4\times10^{-16}$~kg/m$^3$, yielding the wanted effect. However, it would have to be a special kind of dark matter that was not seen in other non-gravitational processes. Dark matter in the form of a spherical halo of a degenerate gas of heavy neutrinos around the Sun (\cite{MunyanezaViollier99}) and `mirror matter' (\cite{FootVolkas01}) have also been discussed.

\noindent{\it String theory and higher-dimensional models.} \cite*{orfeu04} have shown that a generic scalar field cannot explain $a_P$; on the other hand, a non-uniformly-coupled scalar could produce the wanted effect. Although brane-world models with large extra dimensions may offer a richer phenomenology than standard scalar-tensor theories, it seems difficult to find a convincing explanation for the Pioneer anomaly (\cite{orfeu_jorge05}).

\noindent{\it Further ideas.} These include Yuka\-wa-like or higher order corrections to the Newtonian potential (\cite{pioprd}); a scalar-tensor extension to the standard gravitational model (\cite{NovatiCapozzielloLambiase00}); Newtonian gravity as a long wavelength excitation of a scalar condensate inducing electroweak symmetry breaking (\cite{ConsoliSiringo99}); interaction of the spacecraft with a long-range scalar field, unrelated to gravity, determined by an external source term proportional to the Newtonian potential (\cite{MbelekLachieze-Rey99}). In addition, there were suggestions based in flavor oscillations of neutrinos in the Brans-Dicke theory (\cite{Capozzielloetal01}); a theory of conformal gravity with dynamical mass generation, including the Higgs scalar (\cite{WoodMoreau01}). These models are quite interesting, but certainly need more consideration to be regarded as a viable explanation.

\subsection{Search for independent confirmation}
\label{sec:experiemnts}

Attempts to verify the anomaly using other spacecraft proved 
disappointing. This is because the Voyager, Galileo, Ulysses, and
Cassini  spacecraft navigation data all have their own individual difficulties for use in an independent test of the anomaly. In addition, many of the deep space missions that are currently being considered either may not provide the needed navigational accuracy and trajectory stability sensitive to accelerations of under $10^{-10}$~m/c$^2$ (e.g., NASA New Horizons mission) or else they will have significant on-board systematics that mask the anomaly (e.g., JIMO -- Jupiter Icy Moons Orbiter). 

To enable a clean test of the anomaly there is also a requirement 
to have an escape hyperbolic trajectory. This makes a number of other missions (i.e., LISA -- the Laser Interferometric Space Antenna, STEP -- Satellite Test of Equivalence Principle,  etc.) less able to directly test the anomalous acceleration.  Although these missions all have excellent scientific goals and technologies, nevertheless, their orbits lend them a less advantageous position to conduct a precise test of the detected anomaly. 

A number of alternative ground-based verifications of the anomaly have also been considered; for example, using Very Long Baseline Interferometry (VLBI) astrometric observations.  However, the trajectories of spacecraft like the Pioneers, with small proper motions in the sky, make it presently impossible to use VLBI in accurately isolating an anomalous sunward acceleration of the size of $a_P$.

To summarize, the origin of this anomaly remains unclear. \cite*{stanford} advocated a program to study the Pioneer anomaly which effectively has three phases:
\begin{itemize}
\item[i)] Analysis of the entire set of existing Pioneer data, obtained from launch to the last useful data received from Pioneer 10 in April 2002.  This data could yield critical new information about the anomaly (\cite{mex,stanford}).   If the anomaly is confirmed, 
\item[ii)] Development of an instrument, to be carried on another deep space mission, to provide an independent confirmation for the anomaly.  If further confirmed, 
\item[iii)] Development of a dedicated deep-space experiment to explore
the Pioneer anomaly with an accuracy 
for acceleration resolution at the level of $10^{-12}$~m/s$^2$ in the extremely low frequency (or nearly DC) range. 
\end{itemize}
 
An effort to retrieve the early data, existing on obsolete-format magnetic tapes, and transfer it to modern DVDs is being initiated at JPL.  In August 2005 this transfer will be complete, making the entire data record available for the community.  A preliminary analysis of the entire data confirms the previous results on the anomaly and supports its extravehicular nature, thus necessitating experimental work to find its origin. 

Recently there was a significant progress concerning the second phase (\cite{pio-mission,mex,cospar,andreas_pluto,andreas}). This work demonstrated a limited use of an instrumental package to test the Pioneer anomaly emphasizing the need for a dedicated mission to explore the Pioneer anomaly. Finally, the work on phase iii) has also recently been initiated (\cite{enigma,cosmic,mex,stanford}). 

In the following Section we discuss the work needed to experimentally find the origin of the Pioneer anomaly.


\section{A Mission to Explore the Pioneer Anomaly}
\label{missions}

Recent (2004-2005) mission studies have identified two options: i) an experiment on a major mission to deep space capable of reaching acceleration sensitivity similar to that demonstrated by the Pioneers. This option would have a major impact on spacecraft and mission designs with questionable improvement in measuring $a_P$.  On the other hand ii) a dedicated mission to explore the Pioneer anomaly that offers full characterization of the anomaly.  These studies led to a realization that, in order to perform an independent and clear test of the Pioneer anomaly, one needs a new dedicated experiment.  Furthermore, there is an evident `win-win' situation with both outcomes – standard and new physics, are clearly being very important.  Thus, if the anomaly finds its origin in standard physics, such an explanation will be important for solar system physics, astrophysics, and also for advanced high-accuracy navigation.  However, there is a possibility for discovering new physics with truly amazing opportunities.  

From the above, it is clear that a dedicated mission is both scientifically and technologically attractive. Nevertheless, below we describe both possibilities. 

\begin{figure*}[ht!]
 \begin{center}
\noindent    
\psfig{figure=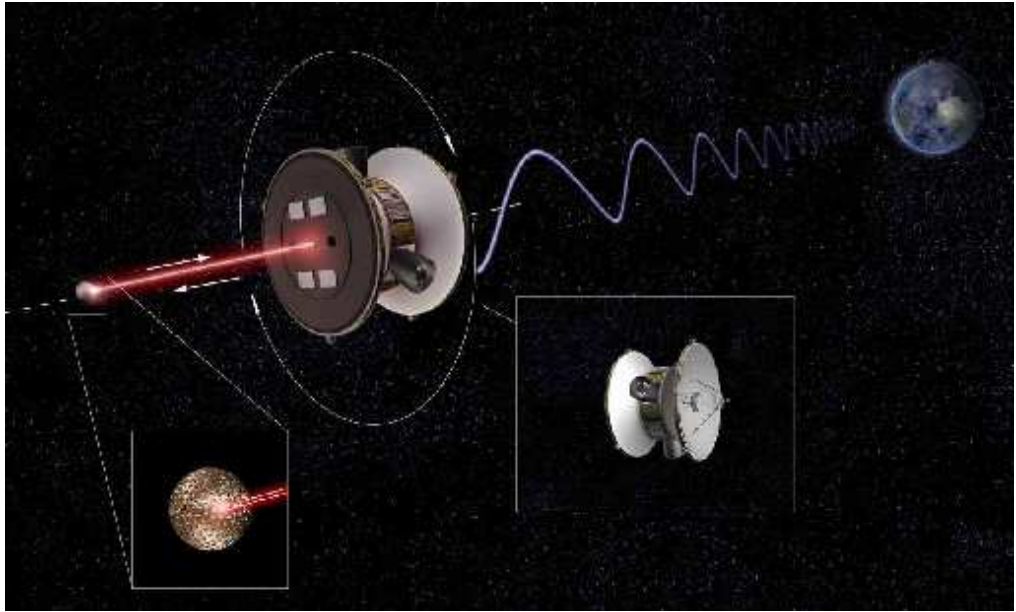,angle=-90,width=13.4cm}
\end{center}
  \caption{A drawing for the measurement concept chosen of the Pioneer Anomaly Explorer.  The formation-flying approach relies on actively controlled spacecraft and a set of passive test-masses. The main objective is to accurately determine the heliocentric motion of the test-mass by utilizing the 2-step tracking needed for common-mode noise rejection purposes.  Trajectory of the spacecraft will be determined based on the standard methods of radio-tracking; while the motion of the test-mass relative to the spacecraft will be established by laser ranging technology.  The test mass is at an environmentally quiet distance from the craft, $\geq$250~m. With occasional maneuvers to maintain formation, the concept establishes a flexible craft-to-testmass formation.
 \label{fig:pae-concept}}
\end{figure*} 


\subsection{Mission Objectives and Requirements}

Experience with studying the Pioneer spacecraft and our current understanding of the Pioneer anomaly lead to the following set of science objectives and technological goals for a dedicated mission to explore it: 
\vskip 5pt
\noindent{\it Scientific Objectives: }
\begin{itemize}
\item Investigate the origin of the Pioneer anomaly with an improvement by a factor of 1,000; 
\item Improve spatial, temporal, and directional resolution; 
\item Identify and measure all possible disturbing and competing effects;
\item Test Newtonian gravity potential at large distances;
\item Discriminate amongst candidate theories explaining $a_P$;
\item Study the deep-space environment in the outer solar system; 
\item Improve limits on the extremely low-frequency gravitational radiation.
\end{itemize}
\noindent{\it Technological Goals} -- to develop the following:
\begin{itemize}
\item methods for precise spacecraft navigation \& attitude control (needed for all future interplanetary missions);
\item drag-free technologies operating at extremely low-frequ\-encies (needed for next generation of GW missions); 
\item  fast orbit transfer scenarios for deep-space access, namely propulsion concepts (including solar sails) and power management at large heliocentric distances (including the use of RTGs);
\item  advanced on-board environmental sensors.
\end{itemize}

The experience gained from the Pioneer spacecraft leads to a creative approach to spacecraft and mission design that responds directly to the set of objectives and goals presented above (\cite{pio-mission,mex,cospar,stanford}). In particular, this experience translates in the following design requirements for the new mission, that are characterized by
\vskip 5pt 
\noindent {\it Navigation and Attitude Control:} 
\begin{itemize}
\item Spin-stabilized spacecraft;  
\item 3-D acceleration sensitivity $\sim 10^{-12}$~m/s$^2$, in very low frequency or DC range;  
\item Propulsion system with precisely calibrated thrusters, propellant lines \& fuel gauges with real-time control;
\item X- and Ka-band with significant dual-band tracking;  
\item Data types: Doppler, range, $\Delta$DOR, and VLBI.
\end{itemize}
{\it Thermal design:}
\begin{itemize}
\item Entire spacecraft is heat-balanced \& heat-symmetric;
\item Active control of all heat dissipation within \& outward;
\item Knowledge of 3D vector of thermal recoil force;
\item Optical surfaces with understood ageing properties;
\item On-board power -- the use of RTGs;
\item Must provide thermal and inertial balance \& stability.
\end{itemize}
{\it Mission Design:}
\begin{itemize}
\item Range of heliocentric distances of interest 25-45 AU;
\item Hyperbolic escape trajectory beyond 15 AU; 
\item Fast orbit transfer with a velocity of $>$ 5 AU / year. 
\end{itemize}

Most of the technology is readily available and could lead to rapid mission design and components fabrication. 

\subsection{A Pioneer Instrument Package}

A way to test for the anomaly would be to fly an instrumental package on a mission heading to the outer regions of the solar system. The primary goal here would be to provide an independent experimental confirmation of the anomaly.  One can conceive of an instrument placed on a major mission to deep space.  The instrument would need to be able to compensate for systematic effects to an accuracy below the level of $10^{-10}$~m/s$^2$.  Another concept would be a simple autonomous probe that could be jettisoned from the main vehicle, such as InterStellar Probe, presumably further out than at least the orbit of Jupiter or Saturn. The probe would then be navigated from the ground yielding a navigational accuracy below the level of $10^{-10}$~m/s$^2$.  The data collected could provide an independent experimental verification of the anomaly's existence.

Such an instrumental payload could, in principle, provide a significant information on the anomalous behavior of the spacecraft; however, in order to explore the anomaly to a greater accuracy of $\sim10^{-12}$~m/s$^2$ one needs a dedicated mission, as discussed in the following chapter.

\subsection{A Dedicated Mission Concept}

Advantages of a dedicated concept include demonstration of new technologies and capabilities, especially in developing technologies for a low disturbance craft, advanced thermal design, formation-flying, accurate navigation and attitude control, etc. It also enables synergies with other science and technology, namely solar system studies (including plasma, dust distributions), Kuiper belt, GWs, heliopause, etc. The goal here would be to explore the anomaly at the $10^{-12}$~m/s$^2$ level in the near DC frequency range and, in so doing, develop technologies critical for future deep-space navigation and attitude control. 

Experience gained from the Pioneer spacecraft leads to a creative approach to spacecraft design.  In particular, we emphasize a precision formation flying as a feasible flight system concept for the proposed mission.  For this architecture, a passive sphere covered with cornercube retroreflectors is laser-ranged from the primary craft. The resulting distance is then combined with the distance between the Earth to the primary craft, determined with radio-metric methods (see Fig.\ref{fig:pae-concept}). Not only will this design allow for the most accurate orbit determination ever, it will also lead to the development of optical navigation, communication, and accurate formation flying technologies.  This mission could also benefit from improvements in low-frequency accelerometers, ultra-stable oscillators, precision star trackers, dust detectors, spectrometers, and real-time autonomous attitude control.

A viable concept would utilize a spacecraft pair capable of flying in a flexible formation.  The main craft would have a precision star-tracker and an accelerometer and would be capable of precise navigation, with disturbances, to a level less than $\sim10^{-10}$~m/s$^2$ in the low-frequency acceleration regime.  Mounted on the front would be a container holding a probe - a spherical test mass covered with cornercubes.  Once the configuration is on its solar system escape trajectory and will undergo no further navigation maneuvers, and is at a heliocentric distance of $\sim5-20$ AU, the test mass would be released from the primary craft.  (This concept is essentially a version of a disturbance-compensation system with a test mass being outside of the spacecraft.)  The probe will be passively laser-ranged from the primary craft with the latter having enough $\Delta$V to maneuver with respect to the probe, if needed.  The distance from the Earth to the primary would be determined with either standard radio-metric methods operating at Ka-band or with optical communication.  Note that any dynamical noise at the primary would be a common mode contribution to the Earth-primary and primary-probe distances. This design satisfies the primary objective, which would be accomplished by the two-staged accurate navigation of the probe with sensitivity down to the $10^{-12}$~m/s$^2$ level in the DC of extremely low frequency bandwidth.  

Since the small forces affecting the motion of a craft from four possible directions, all having entirely different characters (i.e. sunward, Earth-pointed, and along the velocity vector or spin-axis (\cite{pio-mission,stanford})), it is clear that an antenna with a highly pointed radiation pattern and star-sensors will create even better conditions for resolving the true direction of the anomaly, when compared to standard navigation techniques. On a craft with these additional capabilities, all on-board systematics will become a common mode factor contributing to all the attitude sensors and antennas. The combination of all the attitude measurements will enable one to clearly separate the effects of the on-board systematics referenced to the direction towards the Sun. 

To enable fast orbital transfer to distances greater than 20 AU, hyperbolic escape trajectories enabled by solar sail propulsion technology are considered as an attractive candidate.  Among other options is standard chemical rocket and nuclear electric propulsion, as  was successfully demonstrated recently.  The proposed combination of a formation-flying flight system aided by solar sail propulsion for fast trajectory transfer, leads to a technology combination that will benefit many astronomy and fundamental physics missions in the future.

\section{Conclusion}
\label{conclude}

We discussed a joint European-US proposal to explore the anomalous signal of the Pioneer spacecraft -- a mission proposed as a Theme for ESA's Cosmic Vision 2015-2025 (\cite{cosmic}). This mission is designed to determine the origin of the discovered anomaly and to characterize its properties to an accuracy of at least three orders of magnitude below its measured value.  This mission could be an excellent opportunity to demonstrate new technologies for spacecraft design, in-space propulsion, on-board power, deep space communication, and attitude control.  They all could find their way into many fundamental physics and space exploration applications in the future.

The existence of the Pioneer anomaly is no longer in doubt.  Further, after much understandable hesitancy, a steadily growing part of the community has concluded that the anomaly should be subject to 
interpretation.  Our program presents an ordered approach to doing this.  The results would be win-win; improved navigational protocols for deep space at the least, exciting new physics at the best.  Finally, a strong international collaboration would be an additional outcome of the proposed program of understanding the Pioneer anomaly.

A mission to explore the Pioneer Anomaly is motivated by the desire to better understand the laws of fundamental physics as they affect dynamics in the solar system.  This objective is supported by a technology roadmap for development of critical infrastructure elements and accompanied by ordered approach to accomplish this goal.  Such a combination, aided with the political will of the European Space Agency, may not only result in a unique space experiment, but may also become a critical turning point for European endeavors in the deep-space exploration. 

\begin{acknowledgements}

The work of SGT and JDA  was carried out at the Jet Propulsion Laboratory, California Institute of Technology, under a contract with the National Aeronautics and Space Administration. 
Alexandre Szames contributed with graphic design.

\end{acknowledgements}



\end{document}